\documentclass[%
 reprint,
showpacs,
 amsmath,amssymb,
 aps,
]{revtex4-1}

\usepackage{graphicx}% Include figure files
\usepackage{dcolumn}% Align table columns on decimal point
\usepackage{subcaption}
\usepackage{epstopdf}

\usepackage{bm}% bold math
\begin{document}

%\begin{frontmatter}

%\title{Sound Propagation in lined ducts with obstacles using Mild-Slope formulation. \tnoteref{mytitlenote}}
\title{Acoustic Propagation in lined ducts with varying cross-section using a Mild-Slope approximation. }

\author{Maaz Farooqui, Yves Aur\'egan  and Vincent Pagneux}
\address{Laboratoire d'Acoustique de l'Universit\'e du Maine, UMR CNRS 6613
Av. O Messiaen, F-72085 LE MANS Cedex 9, France}

%% or include affiliations in footnotes:

\begin{abstract}
 A modelling of low-frequency sound propagation in slowly varying ducts with smoothly varying lining is  proposed leading
 to an acoustic mild-slope equation analogue to the with mild-slope equation for water waves.
   This simple 1D Mild Slope Equation is derived by direct application of the  Galerkin method. 
   It is shown that the acoustic mild-slope equation can serve as a good alternative to computationally expensive Helmholtz equations to solve such kind of problem. The results from this equation agrees well with FEM based solutions of Helmholtz equation. 
\end{abstract}


%\end{frontmatter}

%\linenumbers

\maketitle

\section{Introduction}

The theory of sound propagation in straight ducts with constant impedance type boundary conditions
and a homogeneous (stationary) medium is classical and well-established \cite{morse1968theoretical}.
 In certain applications the assumption of a straight duct and constant impedance is not valid, and it is therefore of practical interest to consider sound transmission through lined
ducts of varying cross-section. 
A well-known approximation to acoustic propagation in smoothly varying hard wall duct is given by the
horn equation \cite{morse1968theoretical,webster1919acoustical}, which still is a subject of  interest for several applications \cite{rienstra2005webster,gupta2015design}. 
When the smooth variation in duct cross-section is coupled with variable impedance, 
computationally expensive numerical methods remains the preferred choice. 
Indeed, although acoustic propagation in slowly varying ducts has been widely investigated for many decades \cite{rienstra2003sound,
ovenden2004mode,rienstra2001cut,peake2001acoustic,
brambley2008sound,nayfeh1973acoustic,
rienstra2001numerical}, efficient approximate equation to solve this problem does not exist yet. 

 In the context of water waves, to investigate the effect of mild slope water-beds on propagation,   a classical tool is
the  Mild-Slope Equation (MSE) \cite{JUNG2008835,chamberlain1993wave,berkhoff1976mathematical,booij1983note} .
  An improved version of this formulation called Modified Mild-Slope Equation (MMSE) \cite{chamberlain1995modified,liu2013series,suh2005long,liu2012analytic} was later used to study variety of depth variations such as smooth beds \cite{liu2013series} and even circular bowl pits \cite{suh2005long}. 
  In the following we will use the similarity between water waves on varying bathymetry and acoustic propagation in smoothly varying lined ducts.    
  By this simple analogy an  Acoustic Mild Slope Equation (AMSE) is derived by direct application of the classical Galerkin method based on vertical integration used by  Berkhoff \cite{ berkhoff1976mathematical}. An improved  form of this equation, termed 
  Modified Acoustic Mild Slope Equation (MAMSE) is also derived. The approximate results obtained from the AMSE and the MAMSE are compared with numerical solutions of Helmholtz Equation and the results are in close agreement. This 1D approximation equation promises easier implementation and computationally efficient way to solve problems of lined ducts with or without slowly varying cross-sections.

\begin{figure}[!htb]
     \centering
      	  \includegraphics[width=.85\columnwidth]{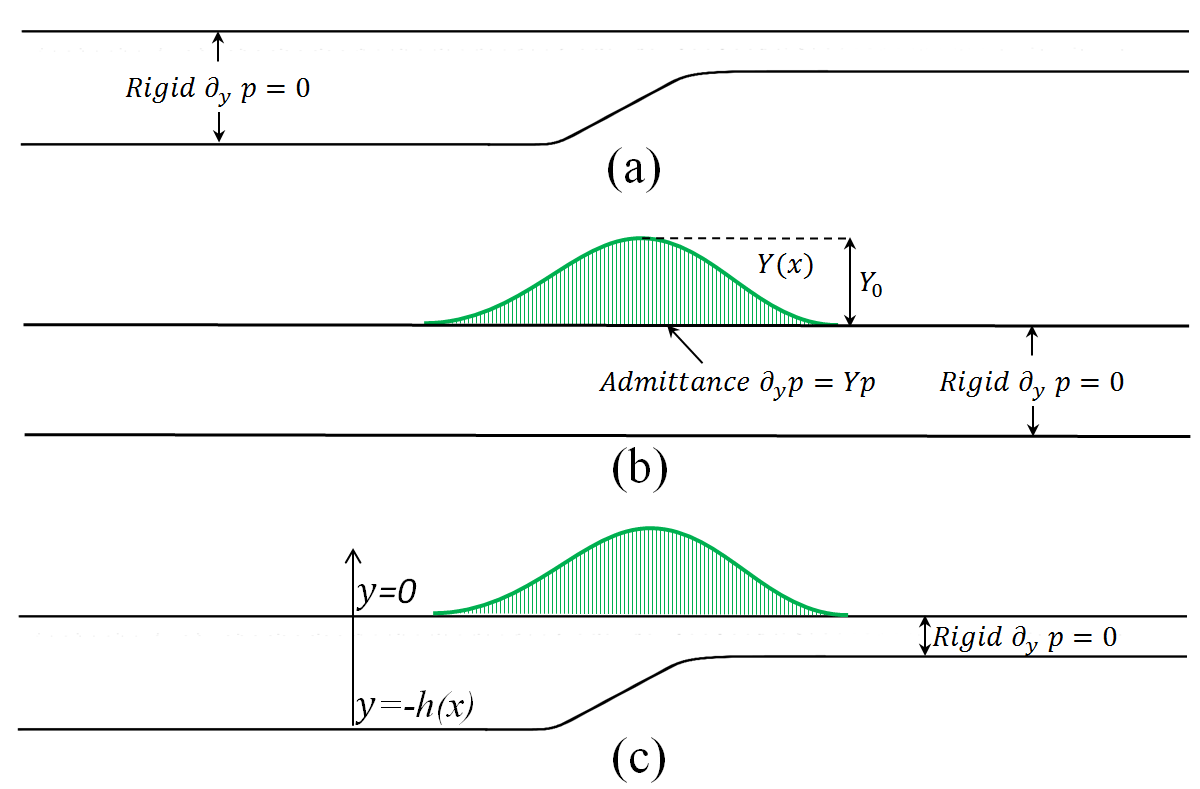}
%                          \caption{Lined duct}\label{fig:1.1}

%                          \caption{Lined duct with varying cross-section}\label{fig:1.2}
                          \caption{Schematic under investigation, (a) Smoothly varying rigid duct, (b) A lined duct and (c) Lined duct with smoothly varying cross-section.}\label{fig:1}
\end{figure}

\section{Theory}

  We consider the sound propagation in a 2D channel, see (Fig. 1(b)). The lower wall is rigid while the upper wall is compliant and described by a varying admittance $Y(x)$. When the distances are non-dimensioned by the height of the channel $H$, the Helmholtz equation, governing the propagation of the acoustic pressure $p$, is:
\begin{equation}
\label{eq:1}
\Delta p + {k^2} p = 0
\end{equation}
  where  $k =\omega H /c_0$  is the reduced frequency, $\omega$ is the frequency and $c_0$ is the sound velocity. 
The boundary conditions are $\partial_n p=0$ for $y=-h(x)$ and $\partial_y p=Y p$ for $y=0$. 

For a uniform admittance Y, a solution of the form $p = A \cosh(\alpha (y)) \exp(\mathrm{i}(-\omega t + \beta x))$ is searched where $\alpha^2 = \beta^2-k^2$ and this leads to the dispersion relation:
\begin{equation}
\label{eq:2}
Y = {\alpha}\: \text{tanh}({\alpha{h}})
\end{equation}
  In the following this equation will be also used for varying $Y$ and $h$ to get a local value $\alpha(x)$. 
With some  manipulations it is possible to obtain a simple 1D equation that explains  the three cases depicted in Fig. 1.  
Indeed, following the same lines as in \cite{chamberlain1995modified}, we obtain
a 1D acoustic mild slope equation (AMSE) that is written as
 \begin{equation}
\label{eq:3}
{\nabla _x}. {\left( {{u_0(x)}{\nabla _x}{p_o}} \right)}  + \left(( {{\alpha ^2+k^2)}{u_0(x)}} \right){p_o} = 0
\end{equation}
 Where
  \begin{equation}
\label{eq:5}
 \begin{gathered}
{u_0}(x) =  \int\limits_{-h(x)}^0 {{w_0}^2} dy 
 =\frac{1}{{2\alpha^2}}\left[ {h(x)(\alpha^2-Y^2) + Y} \right]\\ = \frac{1}{{2\alpha}}\left[ \frac{(Yh(x))'}{(\alpha{h(x)})'} \right]
\end{gathered} 
 \end{equation} 
 $$p = p_0(x) w_0(y;x)$$
 and 
  $$w_0(y;x)=\text{sech}{(\alpha{h(x)})}\cosh(\alpha(x)(y+h(x))).$$
 Here, the wavenumber $\alpha$ is the real, positive root of the  local dispersion relation.

Since its derivation, the MSE has proved to be a very powerful tool to model  linear water wave propagation, because in addition to providing information about both refraction and diffraction effects also it was significantly accurate for short water waves as well as long water waves. 
%In analogy to acoustics, it was expected to perform well for both low and high frequency sound. 
%The importance of the MSE is evident from the several hundred research articles but none of them dealt with its application in acoustic domain. 
For acoustics, the AMSE that we propose will be limited to low frequencies where only one mode is propagating in the waveguide.
%Taking into account the $u_0(x)$ term in Eqn. (\ref{eq:3}), it can be seen that the derivatives $(Y(x)h(x))'$ and/or  $(\alpha{h(x)})'$, have a very important role to play in this equation. 
Besides, without lining (for hard wall ducts $Y=0$), the AMSE should be able to recover the classical horn equation \cite{rienstra2005webster}.
Indeed, it is the case: taking the limit of small $Y$ going to zero, the dispersion relation (Eqn. \ref{eq:2}) becomes, $Y=\alpha^2h(x)$, for $\alpha  \ll {\text{1}}$, which implies, $u_0(x)=h(x)$. Putting this value in Eqn. (\ref{eq:3}) we obtain 
 \begin{equation}
\label{eq:8}
  p(x)''  +  \frac{h(x)'}{h(x)}p(x)' + k^2 {p(x)} = 0 
\end{equation}
that shows that the AMSE transforms to the classical acoustic horn equation for the hard wall case ($Y \rightarrow 0$).
%Similarly,  $Y=\alpha$, for $\alpha  \gg {\text{1}}$, which implies, $u_0(x)=1/(2\alpha)$. In this case, the MSE will cease to depend on the be 
 
% \begin{equation}
%\label{eq:8.5}
%  p(x)''  +  \frac{h(x)'}{h(x)}p(x)' + ({{\alpha ^2+k^2)}} {p(x)} = 0 
%\end{equation}

   %This equations turns into Webster's equation \cite{rienstra2005webster} for $\alpha=0$. 
   
   There had been an extension to MSE which is called the modified mild-slope equation (MMSE) initially derived
by Chamberlain and Porter \cite{chamberlain1995modified}, in which both the obstacle curvature term related to $\Delta^2h$ and the slope-squared term
related to $(\Delta{h})^2$ are added into the traditional AMSE (Eqn. \ref{eq:3}). In the water wave problems, it is shown that this equation is capable of describing known scattering properties of singly and doubly periodic ripple beds, for which the mild-slope equation fails. 
Retaining a $r(x)$ term  we get the Modified Acoustic Mild Slope Equation (MAMSE).
 \begin{equation}
\label{eq:6}
{\nabla _x}. {\left( {{u_0}{\nabla _x}{p_o}} \right)}  + \left(( {{\alpha ^2+k^2)}{u_0} + r(x)} \right){p_o} = 0
\end{equation}
 where
  \begin{equation}
\label{eq:7}
 \begin{gathered}
  r(x) = \int_{-h(x)}^0 {{w_0}{\nabla _x}^2{w_0}dy} \,\,  %\text{where}, w_0=\text{sech}{(\alpha{h})}\cosh(\alpha(y+h))
\end{gathered} 
 \end{equation}

%\subsection{Frequency dependence}

\section{Results and Discussions}

In this section, the results from the AMSE and the MAMSE are discussed and are compared with solutions of the Helmholtz Equation using Finite Element Method (FEM) based COMSOL.
  
 Among others, a practical realization of the admittance can be done by using small closed tubes of variable lengths $b(x)$ perpendicular to the upper wall. Considering lossless tubes, the admittance can be written as:
\begin{equation}
\label{eq:9}
Y (x,{k}) = k\: \tan(k \: b).
\end{equation}
In what follows, a variation of the tube length $b(x)$ is selected as
\begin{equation}
\label{eq:10}
b(x) = \frac{b_0}{2}(1+cos(\frac{\pi{x}}{{{\text{L}}_{{\text{imp}}}}}))
\end{equation}
where  ${{\text{L}}_{{\text{imp}}}}$ and $b_0$ are the lengths which characterizes the distance and max height over which the admittance varies.
Besides, the shape of the obstacle will be given by
\begin{equation}
\label{eq:10}
h(x) = \frac{H_{obs}}{2}(1+cos(\frac{\pi{x}}{{{\text{L}}_{{\text{obs}}}}}))
\end{equation}
\begin{figure}[ht!]
     \centering
      	  \includegraphics[width=.6\columnwidth]{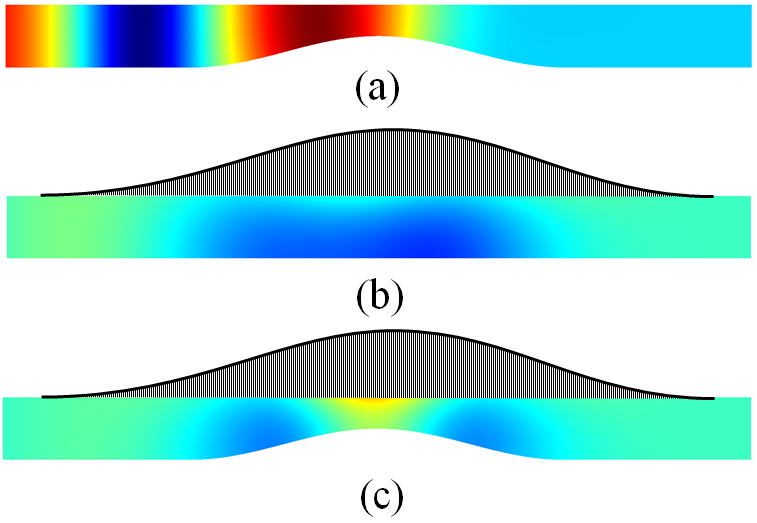}
                          \caption{Real Pressure field ($k=0.2\pi$) for three configurations, (a) Smoothly varying rigid duct, (b) A lined duct and (c) Lined duct with smoothly varying cross-section.}\label{fig:2}
\end{figure}

\subsection{Mild Slope - AMSE}

A number of numerical experiments were carried out in order to check the
1D mild-slope equation against the 2D Helmholtz equation. A reference solution will be given by the numerical FEM computation
form the Helmholtz equation,
%is a well known approximation for this phenomenon and 
%is discretized using Finite Element
%Method, due to its flexibility regarding the representation of curved boundaries. 
where triangular mesh is chosen as finite elements in the computational domain. 
The Mild-slope equation (Eqn. \ref{eq:3}) is discretised using fourth order Runge-Kutta method. In order to minimize the computational time, an explicit wave number formulae is used \cite{explicit2018}. The numerical accuracy of both models was determined by varying the
mesh size. Due to one-dimensionality the mild-slope approximation required lower number of dimensions, a smaller mesh size and 
it is therefore computationally much cheaper. 
In the test cases presented, the obstacle heights are chosen such that the horn equation validity \cite{rienstra2005webster} of ($kh\ll{1}$, $h'\ll{1}$) is verified for the hard cases . 
   \begin{figure}[!htb]
     \centering \includegraphics[width=.8\columnwidth]{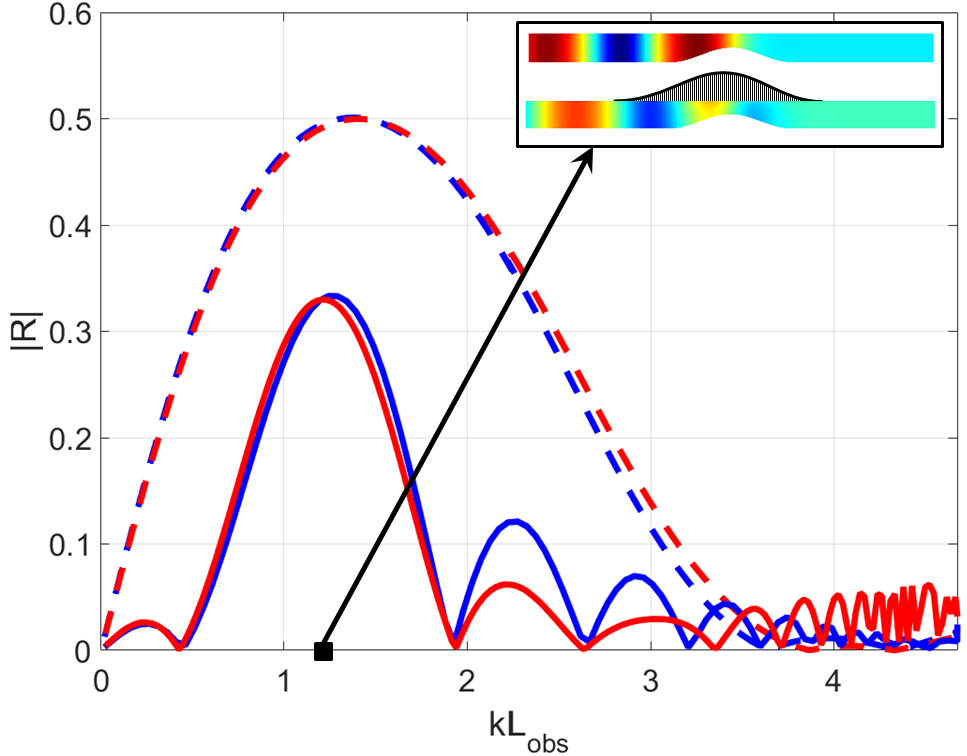}
                   \caption{Absolute reflection $|\text{R}|$ from AMSE solutions for Hard duct red (dashed), lined duct red (solid), FEM solutions for Hard duct blue (dashed) and for lined duct blue (solid) for the parameters ${{\text{L}}_{{\text{obs}}}}=3,  {{\text{L}}_{{\text{imp}}}}=5, H_{obs}=0.5, H_{obs}'=0.26178, Y=0.4645, Y'=0.1432$ (Inset at ${\text{k}}{{\text{L}}_{{\text{obs}}}}=1.28$) }\label{fig:3}
                      \end{figure}                                 
% \begin{figure}[!htb]
%     \centering
%      % \begin{minipage}[b]{0.49\textwidth}
%       \begin{subfigure}[b]{\linewidth}
%       \centering
%	  \includegraphics[width=.75\columnwidth]{Figures/y_53_05n}
%         \caption{$(Yh)'/{\text{k}}$}\label{subfig-1:u}
%       \end{subfigure}\\[\baselineskip]
%      %%    \end{minipage}
%      %   \begin{minipage}[b]{0.49\textwidth}
%       \begin{subfigure}[b]{\linewidth}
%       \centering
%	  \includegraphics[width=.75\columnwidth]{Figures/yhdashn}
%         \caption{$(Yh)'$}\label{subfig-1:u}
%       \end{subfigure}\\[\baselineskip]
%      %    \end{minipage}
%                   \caption{(a) Variation of quantity $(Yh)'/{\text{k}}$ w.r.t ${\text{k}{{\text{L}_\text{obs}}}}$ and longitudinal dimension x(m) for the lined duct with obstacle in Fig \ref{fig:2.5}.(b) Variation of quantity $(Yh)'$ w.r.t $\text{k}$ and ${{\text{L}_\text{obs}}}$ for the lined ducts with $h=0.5$, discussed in majority of results in this work. }\label{fig:3}
%                      \end{figure}                 

Using FEM, 2D Helmholtz Equation was solved to calculate pressure distributions for the three cases shown in Fig. \ref{fig:2}. Fig. \ref{fig:2}(a), (b) and (c), shows the smoothly-varying hard duct,  smoothly-varying impedance and the combination of the two cases, respectively. Although, Fig. \ref{fig:2}(c) is the perfect demonstration of application of the AMSE (Eqn. \ref{eq:3}), it can also efficiently describe the scattering phenomenon of all these configuations. The comparison of results from AMSE and FEM is shown in Fig. \ref{fig:3}. The results shows the variation of absolute reflection coefficient $|\text{R}|$ with the ${\text{k}}{{\text{L}}_{{\text{obs}}}}$, where ${{\text{L}}_{{\text{obs}}}}$ is the length of the obstacle in the duct, which is also referred as length of variation of the duct cross section. AMSE calculations for hard duct (red (dashed)) with $\alpha=0$ (Webster horn equation), is in close agreement with the FEM computations (blue (dashed)), under the accuracy limits of the horn equation ($kh\ll{1}$, $h'\ll{1}$) \cite{rienstra2005webster}.  As our range of interest lies under the horn equation regime, the values of $Y$ and $Y'$ mentioned in figure captions are their maximum values corresponding $k_{max}=1$. The red (dashed) from AMSE and blue (dashed) from FEM refers to hard duct with an obstacle. They seem to agree well till the Webster limit.  This curve corresponds to smoothly varying rigid duct case as depicted in Fig. \ref{fig:2}(a). The  red (solid) from AMSE solutions and blue (solid) from FEM solutions refers to lined duct with obstacle (Fig. \ref{fig:2}(c)). It can be seen that there is significant mismatch between the two from ${\text{k}}{{\text{L}}_{{\text{obs}}}} \geqslant {\text{2}}$. 
This mismatch would need further insights to be quantified in order to better understand the limits of the AMSE with respect to Helmholtz equation. \\

%It is known from Eqn. (\ref{eq:5}) that, variation of $(Y(x)h(x))'$ decides the dominating effect of $p(x)'$ term in the MSE. Fig. \ref{fig:3}(a) depicts the variation of $(Y(x)h(x))'/k$  with respect to the ${\text{k}}{{\text{L}}_{{\text{obs}}}}$ and the longitudinal dimension x. On comparison of  Fig. \ref{fig:2.5} with  Fig. \ref{fig:3}(a), one gets an idea of the mismatch scaling between the MSE and Helmholtz equation. The lowest values of the quantity $(Y(x)h(x))'/k \ll 1$ in Fig. \ref{fig:3}(a), shown by light green, corresponds to the perfect match. This limit is in addition to limits proposed by Webster($kh\ll{1}$, $h'\ll{1}$). As the value increases, the mismatch begins to appear with significant mismatch appearing as  $|(Y(x)h(x))'/k|\simeq 0.3$. This dependence of the scattering on the $|(Y(x)h(x))'|$ is plotted in Fig. \ref{fig:3}(b) for different ${{\text{L}}_{{\text{obs}}}}$, to quantify the accuracy limit of MSE.\\

\begin{figure}[ht!]
     \centering
    	  \includegraphics[width=.8\columnwidth]{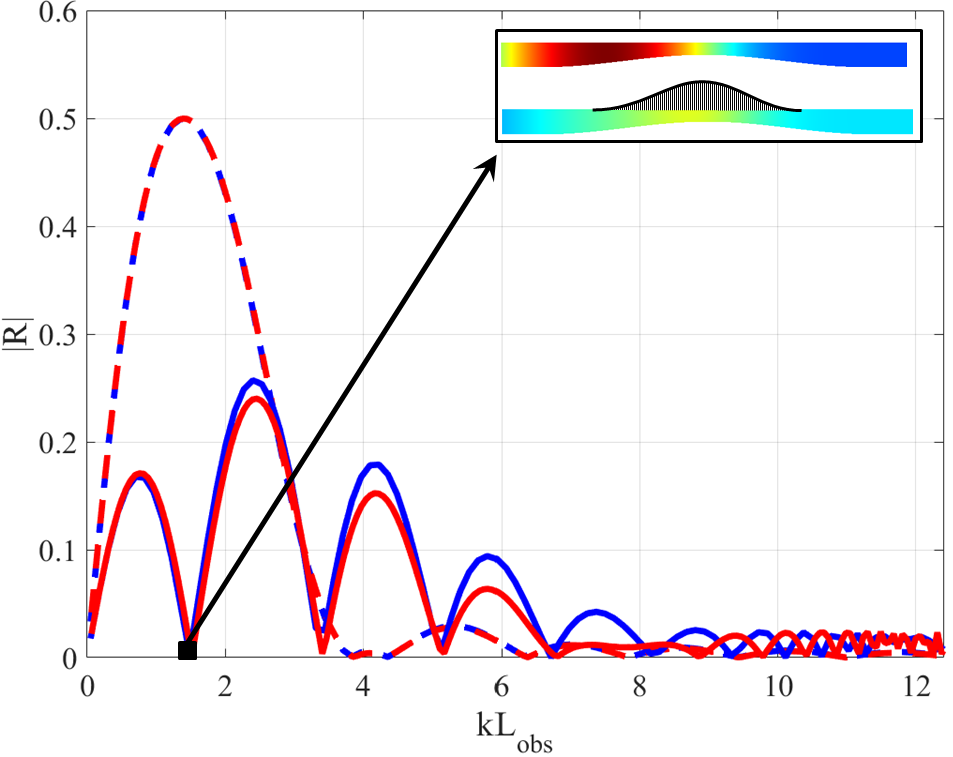}
                          \caption{Absolute reflection $|\text{R}|$ from AMSE solutions for Hard duct red (dashed), lined duct red (solid), FEM solutions for Hard duct, blue (dashed) and for lined duct, blue (solid) for the parameters ${{\text{L}}_{{\text{obs}}}}=8, {{\text{L}}_{{\text{imp}}}}=5, H_{obs}=0.5, H_{obs}'=0.098175, Y=0.4645, Y'=0.1432$ , (Inset at ${\text{k}}{{\text{L}}_{{\text{obs}}}}=1.48$) }\label{fig:4}
                      \end{figure}

\begin{figure}[ht!]
     \centering
    	  \includegraphics[width=.8\columnwidth]{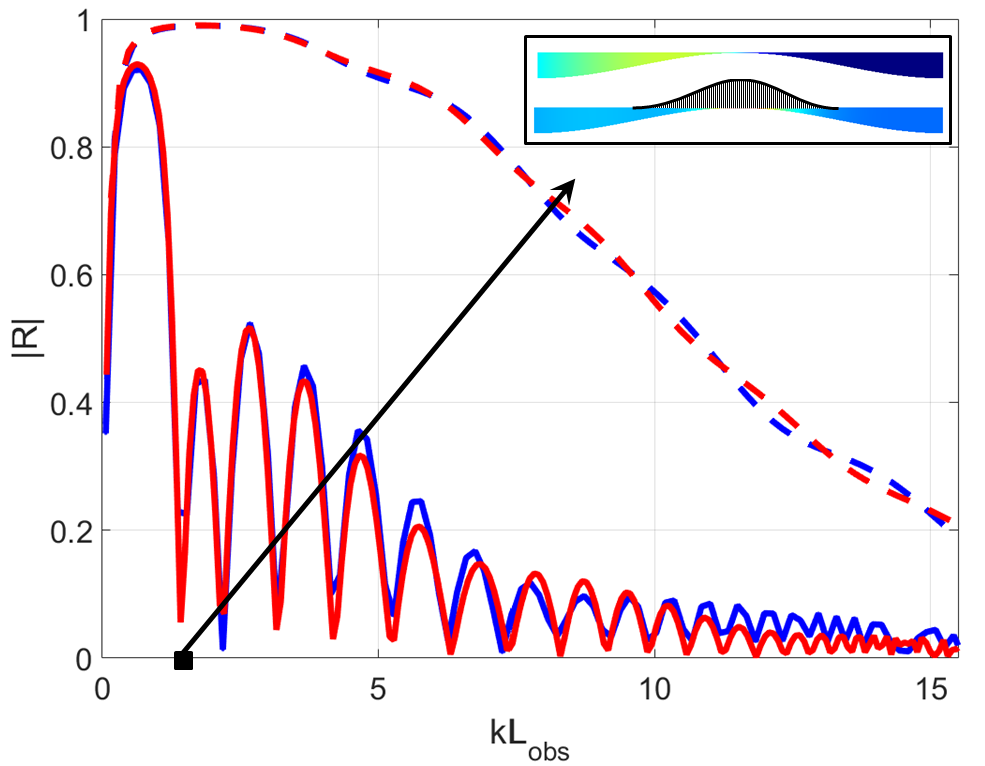}
                          \caption{Absolute reflection $|\text{R}|$ from AMSE solutions for Hard duct red (dashed), lined duct red (solid), FEM solutions for Hard duct blue (dashed) and for lined duct blue (solid) for the parameters $      {{\text{L}}_{{\text{obs}}}}=10, {{\text{L}}_{{\text{imp}}}}=5, H_{obs}=0.98, {\delta}H_{obs}=0.15393, B_{imp}'=0.31416$, (Inset at ${\text{k}}{{\text{L}}_{{\text{obs}}}}=1.45$)  }\label{fig:5}
                      \end{figure}

Fig. \ref{fig:4} and Fig. \ref{fig:5} shows the variation of absolute reflection coefficient $|\text{R}|$ with ${\text{k}}{{\text{L}}_{{\text{obs}}}}$, for ${{\text{L}}_{{\text{obs}}}}=$ 8 and  10, respectively. On comparison with Fig. \ref{fig:3}(b), it can be concluded, that for ${{\text{L}}_{{\text{obs}}}}=$ 8, highly accurate predictions are possible with AMSE, when ${\text{k}}{{\text{L}}_{{\text{obs}}}} \leqslant $ 5. The scaling of Fig. \ref{fig:5} can be obtained re-calculating the result of Fig. \ref{fig:3}(b) with $h=0.98$, which yields ${\text{k}}{{\text{L}}_{{\text{obs}}}} \leqslant $ 10.  In some cases the solutions from AMSE agrees quite well even beyond the Webster limit. The reason to this behavior is extremely low values of $(Y(x)h(x))'$, which aids in extending the proposed limits of AMSE.

\subsection{Modified Mild Slope - MAMSE}

In the domain of water waves, the scattering of water waves by ripples in a horizontal bed  falls outside the scope of the mild-slope equation. $Kirby (1986)$ derived an alternative equation which allowed for a rapidly varying, small-amplitude bedform to be superimposed on a slowly varying component of topography. As explained earlier, the full form of AMSE including the $r(x)$ term is called modified acoustic mild slope equation MAMSE (Eqn. \ref{eq:6}).\\

\begin{figure}[ht!]
     \centering
    	  \includegraphics[width=.8\columnwidth]{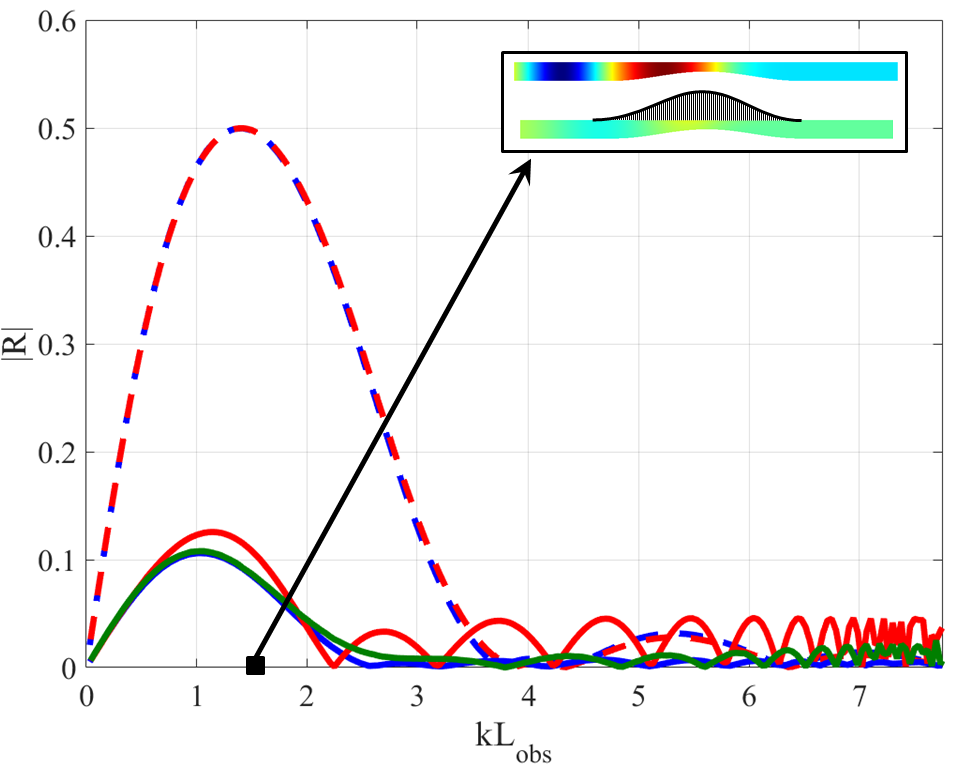}
                          \caption{Absolute reflection $|\text{R}|$ from AMSE solutions for Hard duct red (dashed), lined duct red (solid), FEM solutions for Hard duct blue (dashed) and for lined duct blue (solid) and  MAMSE solutions for lined duct green (solid) for the parameters , ${{\text{L}}_{{\text{obs}}}}=5, {{\text{L}}_{{\text{imp}}}}=5, H_{obs}=0.5, H_{obs}'=0.15708, Y=0.4645, Y'=0.1432$, (Inset at ${\text{k}}{{\text{L}}_{{\text{obs}}}}=1.57$)}\label{fig:6}
                      \end{figure}

Fig. \ref{fig:6} shows the variation of absolute reflection coefficient $|\text{R}|$ with ${\text{k}}{{\text{L}}_{{\text{obs}}}}$, for ${{\text{L}}_{{\text{obs}}}}=$ 5. As evident from the figure, the Mild slope equation (red solid) has a limited accuracy range (${\text{k}}{{\text{L}}_{{\text{obs}}}} \leqslant$ 2). Here, the MAMSE (green solid) surpasses the mild-slope limits and is a much better approximation of the FEM solution (blue curve). Although, application of MAMSE is more difficult and computationally expensive, but as in the case discussed, might serve to extend the AMSE limits for broad frequency range. \\

\begin{figure}[ht!]
     \centering
    	  \includegraphics[width=.8\columnwidth]{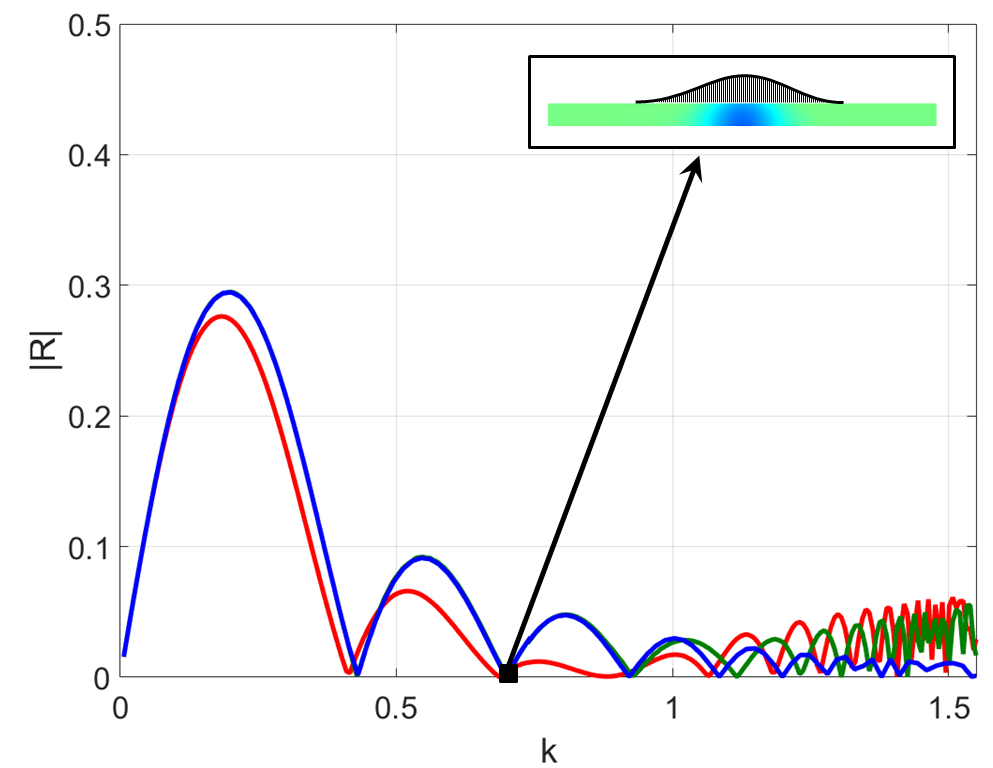}
                          \caption{Absolute reflection $|\text{R}|$ from FEM solutions for straight lined duct blue (solid), AMSE solutions for Non-varying lined duct red (solid), and MAMSE solutions for Non-varying lined duct green (solid) for the parameters , ${{\text{L}}_{{\text{imp}}}}=5,  Y=0.4645, Y'=0.1432$   }\label{fig:7}
                      \end{figure}

In addition to extension of AMSE limits, the inclusion of obstacle curvature term related to $\Delta^2h$ and the slope-squared term
related to $(\Delta{h})^2$ helps in improving the accuracy of the solution. For instance,  Fig. \ref{fig:7} shows  the variation of absolute reflection coefficient $|\text{R}|$ with ${\text{k}}$, for ${{\text{L}}_{{\text{obs}}}}=$ 5. The results are for a straight duct with just a smoothly varying liner. Here, the MAMSE (green solid) is a much better approximation in terms on the broadband prediction as well as accuracy in comparison to Helmholtz equation.             

%Fig. \ref{fig:2} shows real pressure field for three configurations of duct.  The middle lined duct shows the pressure distribution with varying admittance. The bottom duct shows the pressure distribution of varying duct with flush-mounted varying admittance. 2D Helmholtz equation  was solved to calculate these pressure distribution.   The three cases shown can be explained using MSE. The middle case   Fig. \ref{fig:2}(b) is another variation of MSE with $u_0(x)=0$. Fig. \ref{fig:3} shows the variation of absolute reflection coefficient $|\text{R}|$ with the $k{L_{obs}}$, where $L_obs$ is the length of the obstacle in the duct, which is length of variation in the duct cross section.  
%Fig. \ref{fig:2} shows real pressure field for three configurations of duct.  The middle lined duct shows the pressure distribution with varying admittance. The bottom duct shows the pressure distribution of varying duct with flush-mounted varying admittance. 2D Helmholtz equation  was solved to calculate these pressure distribution.   The three cases shown can be explained using MSE. The middle case   Fig. \ref{fig:2}(b) is another variation of MSE with $u_0(x)=0$. Fig. \ref{fig:3} shows the variation of absolute reflection coefficient $|\text{R}|$ with the $k{L_{obs}}$, where $L_obs$ is the length of the obstacle in the duct, which is length of variation in the duct cross section.  
%%

\section{Conclusion}

The mild slope equation is a popular tool to model water wave propagation on a mild-slope bed. In analogy with water waves, a one-dimensional mild-slope formulation is proposed in this work to model low-frequency sound propagation in slowly varying ducts with smoothly varying lining. 
%This is the first time the versatility of  Mild-slope equation is exploited in duct acoustics.   
This approximation is derived by direct application of the classical Galerkins method. 
It is shown that the 1D mild-slope equation is an economical and efficient alternative to computationally expensive and 
complex 2D Helmholtz equations to solve such kind of problem. 
%The results from the mild-slope equation agrees well with FEM based solutions of Helmholtz equation.

%\section*{References}
%\bibliographystyle{elsarticle-num}
%\biboptions{numbers,sort&compress}
\bibliography{archive_WM}

\end{document}